\UseRawInputEncoding
\documentclass[%
 reprint, 
groupedaddress,
 amsmath,amssymb,
 aps,
prb,
]{revtex4-2}
\usepackage{ragged2e}
\usepackage[pagewise,modulo]{lineno}
\usepackage{xcolor}
\usepackage{graphicx}
\usepackage{dcolumn}
\usepackage{bm}
\usepackage{braket}
\usepackage{lipsum}
\usepackage[hidelinks,colorlinks=true,linkcolor=blue,urlcolor=blue,citecolor=blue]{hyperref}

\newif\ifmarked
\markedfalse
\ifmarked
  \newcommand{\new}[1]{{\color{red}#1}}
  \newenvironment{revblock}{\color{red}}{}
  \linenumbers
\else
  \newcommand{\new}[1]{#1}
  \newenvironment{revblock}{}{}
\fi

\begin{document}

\preprint{APS/123-QED}

\title{Domain wall drift in a ferrimagnet induced by a linearly oscillating \\
in-plane magnetic field}

\author{Murod Mirzhalilov}
\thanks{Corresponding author: mirzhalilov.1@osu.edu}

\author{Jacob Freyermuth}

\affiliation{Department of Physics, The Ohio State University, Columbus, OH 43210, USA}

\date{\today}

\begin{abstract}
We investigate ferrimagnetic domain-wall dynamics under linearly oscillating in-plane magnetic fields. Starting from a ferrimagnetic Lagrangian that retains the field-dependent contribution to the kinetic energy, we derive collective-coordinate equations for the domain-wall position and azimuthal angle. This formulation makes explicit the field-induced inertial terms, including Coriolis-, Euler-, and centrifugal-like contributions, and clarifies how they couple translational and azimuthal dynamics. For an oscillating in-plane field supplemented by a weak static out-of-plane bias field, we show that periodic switching of the wall angle can be rectified into a finite average domain-wall velocity. At low drive frequencies, where the wall angle follows the oscillating field, the induced drift is proportional to the drive frequency rather than the field amplitude, \new{and reversing the static field reverses it}. Above a cutoff frequency, \new{which we derive analytically,} the wall angle can no longer follow the drive and the drift is suppressed. Micromagnetic simulations confirm the analytical predictions and demonstrate that the magnitude and direction of the drift can be tuned by the net spin density. These results identify linearly oscillating fields as a mechanism for frequency-controlled domain-wall motion in ferrimagnets and highlight the role of inertial spin dynamics near magnetic compensation.

\end{abstract}

\maketitle

\section{Introduction} 
Chiral magnetic materials have been a major focus of research for spintronics applications in recent years due to their potential use for the design of next-generation data storage devices \cite{Parkin2008, Emori2013,Yang2021ChiralSpintronics}. Particularly, much study has focused on magnetic domain walls, solitons which exist in magnetic materials with easy-axis magnetic anisotropy \cite{Walk74, Thiele73, Alejos2020DWReview}. Designs including domain wall racetracks have been shown to be very promising for efficient and dense memory devices \cite{Parkin2008, Parkin2015, Yang2015}. Additionally, domain walls have been the focus of more fundamental study in an effort to understand how domain walls move when driven by currents and magnetic fields.

In recent years, antiferromagnets have risen in popularity as an alternative to traditional ferromagnets for magnetic device applications, largely due to their lack of stray fields and switching speeds in the THz regime, both clear advantages over ferromagnets \cite{Baltz2018}. The behavior of domain walls in antiferromagnets is fundamentally different than in ferromagnets, with interesting phenomena such as "relativistic" domain wall motion and the lack of a Walker breakdown \cite{Baryakhtar85,Gomonay2016,Shiino16,Tejerina2020}. However, detection and control of spin textures in antiferromagnets is difficult, so this has limited the potential of antiferromagnets from a device design perspective.

Ferrimagnets, magnets with opposing sublattices of unequal strengths, provide the possibility of combining the advantages of both antiferromagnets and ferromagnets \cite{Kim2022NatMater}. Thus, understanding how domain walls move in ferrimagnets is interesting both physically and from a device design perspective. Many recent studies have focused on domain walls in ferrimagnets driven by spin currents, uniform magnetic fields \new{and anisotropy gradients} \cite{Kim2017NatMater, Oh2017,  Kim2020, Yurlov2021, Li2020}. There has also been some study of the motion of domain walls using time-varying magnetic fields \cite{Pan2018, Jin2021, Hardt2025}.

In this paper, we consider the dynamics of the spins in a ferrimagnet in a general, time-dependent external magnetic field. Building on field-theoretic descriptions of antiferromagnetic and ferrimagnetic texture dynamics \cite{Tveten2013, Tveten2016, Tserk2017, Nakane2021}, we retain the field-dependent kinetic contribution and explicitly project the resulting inertial terms onto ferrimagnetic domain-wall collective coordinates. We identify a field-induced Coriolis-like coupling that appears directly from the applied field dependent kinetic energy and couples translational and rotational degrees of freedom. We find that this term is non-zero for in-plane fields and does not vanish in the antiferromagnetic limit. \new{Additionally, we consider a ferrimagnetic domain wall driven by a linearly oscillating in-plane magnetic field together with a weak static out-of-plane field. The oscillating field repeatedly flips the in-plane orientation of the wall, and the static field biases these flips so that the wall drifts in one direction. The average velocity is proportional to the drive frequency and to the net spin density, so the drift vanishes at spin angular momentum compensation. Above a cutoff frequency, which we derive analytically and which grows with the field amplitude, the wall can no longer follow the drive and the drift stops.} We have performed micromagnetic simulations confirming this motion and demonstrating the relationship between the average domain wall velocity and relevant material parameters, \new{such as the net spin density and the strength of the Dzyaloshinskii-Moriya interaction (DMI)}.

\new{Most studies of domain walls driven by time-dependent fields have used rotating fields. In ferromagnets and ferrimagnets, a rotating field drives the wall by synchronizing the wall angle with the field \cite{Yan2009}, and in ferrimagnets the resulting velocity is proportional to the net spin density and the drive frequency \cite{Jin2021}. A bulk DMI adds a further contribution that allows motion even at spin angular momentum compensation \cite{Liu2022}. In antiferromagnets, where the gyrotropic mechanism is absent, rotating fields move the wall with the help of the DMI \cite{Pan2018,Li2020JMMM}. For the linearly oscillating in-plane field considered here, the drift velocity below the cutoff frequency has the same form as the rotating-field result \cite{Jin2021}, so the linear drive keeps the frequency control of rotating fields. It also offers two practical advantages. First, it has no handedness, so the direction of motion is selected by the sign of a weak static field and is reversed simply by reversing that field. Second, it is easier to generate, since a single stripline suffices, whereas a rotating field needs two orthogonal currents a quarter period apart \cite{Bisig2015}. The role of the DMI also differs from the bulk-DMI case \cite{Liu2022}. There, the DMI twists the wall and contributes to the velocity, whereas here the interfacial DMI pins the wall angle. The oscillating field then moves the wall only above a threshold amplitude, while the DMI itself does not enter the drift velocity.}

\new{Linearly oscillating fields combined with static fields have been studied for ferromagnetic and antiferromagnetic walls. In ferromagnets, a static field above the Walker field makes the wall precess, and a weak oscillating field locks this precession, giving a frequency-controlled velocity within a narrow band around the natural precession frequency \cite{Kim2020PRB}. In antiferromagnets, a static field and an oscillating field, both transverse to the easy axis, produce a drift only at discrete resonance frequencies, near 100~GHz for NiO \cite{Chen2018}. The present work concerns a different magnetic system, a ferrimagnetic wall driven by an in-plane oscillating field and a weak out-of-plane static field. The static field is far below the Walker field, so the wall does not precess on its own, and the drift persists from low frequencies up to a cutoff set by the field amplitude. For our parameters, this cutoff lies at a few hundred megahertz. Such frequencies are readily generated, and no tuning to a resonance is needed. The drift is set by the net spin density, which provides an additional control parameter. It can be tuned through composition or temperature \cite{Kim2017NatMater}, and the drift reverses sign across spin angular momentum compensation.}

In Section~\ref{SecII}, we derive the equations of motion for a ferrimagnet in a time-dependent magnetic field. We then introduce a domain-wall ansatz and obtain the corresponding collective-coordinate equations of motion. In Section~\ref{SecIII}, we consider a domain wall driven by a linearly oscillating magnetic field. We analyze the resulting dynamics, explain the physical origin of the induced motion, \new{derive the threshold field amplitude and the cutoff frequency,} and compare the analytical predictions with micromagnetic simulations. The simulations confirm the \new{frequency-driven} drift\new{, its reversal with the static field, and the cutoff frequency}. Finally, in Section~\ref{SecIV}, we summarize our results and discuss possible future directions.

\section{Theory} \label{SecII}

We model a collinear ferrimagnet as two antiferromagnetically coupled sublattices with unequal moments, $\mathbf{M}_i(x,t)=M_{is}\mathbf{m}_i(x,t)$ for $i=1,2$, where $M_{is}$ is the saturation magnetization of sublattice $i$ and $\mathbf{m}_i$ is a unit vector. Each sublattice obeys the Landau--Lifshitz--Gilbert (LLG) equation \cite{LL35,Gilb04,Martinez2019,Jing2022}
\begin{equation}
\label{Eq.LLG}
\frac{\partial \mathbf{M}_i}{\partial t}
= -\gamma_i\,\mathbf{M}_i\times \mathbf{H}_i^{\mathrm{eff}}
+ \frac{\alpha_i}{M_{is}}\,\mathbf{M}_i\times
\frac{\partial \mathbf{M}_i}{\partial t}.
\end{equation}
Here $\gamma_i$ is the gyromagnetic ratio, $\alpha_i$ is the Gilbert damping, and $\mathbf{H}_i^{\mathrm{eff}}=-\delta U/\delta \mathbf{M}_i$ is the effective field derived from the total magnetic energy $U=\int dV \mathcal{U}$ with $\mathcal{U}$ being the magnetic energy density. We neglect inter-sublattice damping for simplicity and consider a thin, quasi-1D strip with perpendicular magnetic anisotropy (PMA) and interfacial DMI.

It is convenient to introduce the net and staggered fields $\mathbf{m}\equiv \mathbf{m}_1+\mathbf{m}_2$ and $\mathbf{n}\equiv(\mathbf{m}_1-\mathbf{m}_2)/2$ \cite{Hals2011,Oh2019,Gomonay2010,Oh2018,Kim2020,Ivanov19}. These fields satisfy $\mathbf{m}\cdot\mathbf{n}=0$ and $|\mathbf{n}|^2+|\mathbf{m}|^2/4=1$. In the exchange-dominated regime, the sublattices are nearly antiparallel, so $|\mathbf{m}|\ll1$ and $|\mathbf{n}|\simeq1$ \cite{Nakane2021,Tveten2016}. The low-energy dynamics is described by the ferrimagnetic Lagrangian density [see Appendix \ref{App:FiMLagrangian}]
\begin{equation}\label{Eq:FiMLagrangian_mn}
\mathcal{L}
= s\,\partial_t\mathbf{n}\cdot(\mathbf{n}\times\mathbf{m})
+ \delta s\,\boldsymbol{\mathcal{A}}(\mathbf{n})\cdot\partial_t\mathbf{n}
- \mathcal{U}(\mathbf{m},\mathbf{n}),
\end{equation}
where $s=(s_1+s_2)/2$ and $\delta s=s_1-s_2$ are the average and uncompensated spin densities, and $\boldsymbol{\nabla}_{n}\times\boldsymbol{\mathcal{A}}(\mathbf{n})=\mathbf{n}$ is the Berry connection on the unit sphere \cite{Tserk2017}. For textures varying along $x$, we take the energy density to contain homogeneous inter-sublattice exchange, spin stiffness, PMA, interfacial DMI, and Zeeman coupling:
\begin{equation}
\begin{aligned}
\label{Eq:EnergyDensity}
\mathcal{U}
= \frac{A_0}{2}\mathbf{m}^2
+ \frac{A}{2}(\partial_x\mathbf{n})^2
- \frac{K_z}{2}n_z^2
\\
- \frac{D}{2}\hat{\mathbf y}\cdot(\mathbf n\times\partial_x\mathbf n)
- \mathbf M\cdot\mathbf H .
\end{aligned}
\end{equation}
Here $A_0>0$ is the homogeneous inter-sublattice exchange, $A>0$ is the spin stiffness, $K_z>0$ is the easy-axis anisotropy, $D$ is the interfacial DMI strength, $\mathbf M$ is the net magnetization, and $\mathbf H$ is the applied magnetic field.

Integrating out $\mathbf{m}$ in the exchange approximation gives the effective Lagrangian density [see Appendix \ref{App:FiMLagrangianField}]
\begin{equation}
\label{Eq:FiMLagrangian}
\mathcal{L}
= \frac{\rho}{2}\bigl(\partial_t\mathbf{n}-\gamma\,\mathbf{H}\times \mathbf{n}\bigr)^2
+ \delta s\,\boldsymbol{\mathcal{A}}(\mathbf{n})\cdot\partial_t\mathbf{n}
- \mathcal{U}(\mathbf{n}),
\end{equation}
where $\rho=s^2/A_0$ is the inertia of the staggered order and $\gamma=(\gamma_1s_1+\gamma_2s_2)/(s_1+s_2)$ is the effective gyromagnetic ratio. The field-dependent kinetic term is often negligible far from compensation \cite{Oh2017}, but it becomes important near magnetic compensation and in the antiferromagnetic limit, where time-dependent magnetic fields can drive the collective dynamics \cite{Tveten2013,Tveten2016}.

Including damping through the Rayleigh function $R=\alpha s|\partial_t\mathbf n|^2$ for $\alpha_1=\alpha_2=\alpha$, we obtain
\begin{equation}
\label{Eq:EqoM}
\begin{aligned}
\rho\,\mathbf{n}\times \Big[
\partial_t^2\mathbf{n}
&-2\gamma\,\mathbf{H}\times \partial_t\mathbf{n}
-\gamma\,\partial_t\mathbf{H}\times \mathbf{n}
+\gamma^2\,\mathbf{H}\times(\mathbf{H}\times\mathbf{n})
\Big]\\
&= \mathbf{n}\times \mathbf{f}_n^{\mathrm{eff}}
+ \delta s\,\partial_t\mathbf{n}
-2\alpha s\,\mathbf{n}\times\partial_t\mathbf{n},
\end{aligned}
\end{equation}
where $\mathbf{f}_n^{\mathrm{eff}}=-\delta U/\delta\mathbf n$. The term proportional to $\partial_t\mathbf H$ corresponds to the reactive magnetic-field force previously discussed in antiferromagnetic collective-coordinate dynamics \cite{Tveten2013,Tveten2016,Nakane2021}. 
The remaining field-dependent terms originate from the covariant kinetic energy $(\partial_t\mathbf n-\gamma\mathbf H\times\mathbf n)^2$ and can be interpreted as Coriolis- and centrifugal-like inertial contributions \cite{Andreev1980,LLMechanics}. Similar Coriolis and centrifugal terms appear in rotating-frame descriptions of antiferromagnetic textures \cite{Tserk2014}; however, in the present case they are induced by the applied magnetic field itself, rather than by transforming to the frame of a rotating domain wall. 
Although these terms are parametrically small in the exchange limit, scaling as $M_sH/A_0\ll1$, their collective-coordinate projection produces a velocity-dependent tensor with the same antisymmetric structure as the conventional gyrotropic coupling.

To describe a domain wall interpolating between $\mathbf n=\pm\hat{\mathbf z}$, we use the rigid collective-coordinate ansatz
\begin{equation}\label{Eq:DWansatz}
\mathbf{n}(x,t)=
\bigl(\cos\Phi\,\mathrm{sech}\tfrac{x-X}{\Delta},\,
\sin\Phi\,\mathrm{sech}\tfrac{x-X}{\Delta},\,
\tanh\tfrac{x-X}{\Delta}\bigr),
\end{equation}
where $X(t)$ is the domain-wall position, $\Phi(t)$ is the azimuthal angle, and $\Delta=\sqrt{A/K_z}$ is the domain-wall width \new{(see Fig.~\ref{fig:schematic})}. We assume $\Delta$ remains approximately constant for the sub-relativistic dynamics considered here \cite{Shiino16,Caretta20}.

Projecting Eq.~\eqref{Eq:EqoM} onto the collective-coordinate modes $\mathbf n\times\partial_{q_j}\mathbf n$, with $q_i=\{X,\Phi\}$ \cite{Tretiakov2008}, gives
\begin{equation}
\label{ForceEqoM}
M_{ji}\ddot q_i+
\left(f_{ji}^{\mathrm{Cor}}-G_{ji}+\frac{\alpha s}{\rho}M_{ji}\right)\dot q_i
=F_j .
\end{equation}
The generalized force is decomposed as
$F_j=F_j^{\mathrm{eff}}+F_j^{\mathrm{Euler}}+F_j^{\mathrm{Centr}}$.
Here $F_j^{\mathrm{eff}}=\int dV\,\partial_{q_j}\mathbf n\cdot\mathbf f_n^{\mathrm{eff}}$ is the effective conservative force,
$F_j^{\mathrm{Euler}}=\rho\gamma\int dV\,\partial_t\mathbf H\cdot(\mathbf n\times\partial_{q_j}\mathbf n)$ is the Euler-like force, and
$F_j^{\mathrm{Centr}}=-\rho\gamma^2\int dV\,(\mathbf H\cdot\mathbf n)(\mathbf H\cdot\partial_{q_j}\mathbf n)$ is the centrifugal-like force.

The velocity-dependent terms in Eq.~\eqref{ForceEqoM} are governed by
$M_{ji}=\rho\int dV\,\partial_{q_j}\mathbf n\cdot\partial_{q_i}\mathbf n$,
$G_{ji}=\delta s\int dV\,\mathbf n\cdot
(\partial_{q_j}\mathbf n\times\partial_{q_i}\mathbf n)$, and
$f_{ji}^{\mathrm{Cor}}=2\rho\gamma\int dV\,\mathbf H\cdot
(\partial_{q_j}\mathbf n\times\partial_{q_i}\mathbf n)$,
which describe the inertial, gyrotropic, and Coriolis-like couplings, respectively.
The Coriolis-like tensor has the same antisymmetric structure as the conventional gyrotropic tensor, but with $\delta s\,\mathbf n$ replaced by $2\rho\gamma\,\mathbf H$.
Unlike the rotating-frame Coriolis term associated with a precessing texture \cite{Tserk2014}, this coupling is induced by the applied magnetic field and is externally tunable through the magnitude and direction of $\mathbf H$.

This distinction becomes important near spin angular momentum compensation. In the compensated antiferromagnetic limit, $\delta s\to0$, the conventional gyrotropic tensor vanishes. The field-induced Coriolis-like tensor, however, can survive at compensation and generate a residual velocity-dependent coupling between $X$ and $\Phi$ when the applied field has an appropriate in-plane component. This coupling is small in the exchange limit, scaling as $M_sH/A_0$, but it suggests a possible route to Walker-breakdown-like dynamics in compensated antiferromagnets or near-compensated ferrimagnets at sufficiently large fields. Away from compensation, the Zeeman contribution contained in $F_j^{\mathrm{eff}}$ dominates the field-driven dynamics, while the Euler and centrifugal terms are smaller by $M_sH/A_0$ and can be neglected to leading order. We use this hierarchy in the analysis below.

\section{Domain Wall Dynamics with a Linearly Oscillating Field} \label{SecIII}

\subsection{Analytical Analysis}

\begin{figure}
\begin{revblock}
\centering
\includegraphics[width=\columnwidth]{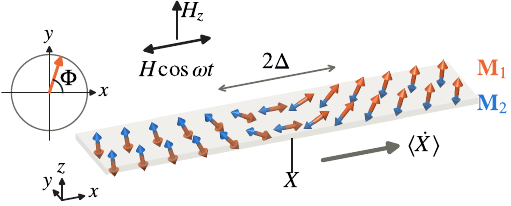}
\caption{Schematic of the model. A domain wall at position $X$ separates the down and up domains of a ferrimagnetic strip with perpendicular anisotropy. At each site, the long orange and short blue arrows show the antiparallel sublattice magnetizations $\mathbf{M}_1$ and $\mathbf{M}_2$, which have unequal magnitudes. The bracket marks the wall region from $X-\Delta$ to $X+\Delta$. The circle on the left is a top view of the $xy$ plane, where the in-plane component of $\mathbf{n}$ at the wall center makes the azimuthal angle $\Phi$ with the $x$ axis. The wall is driven by the in-plane field $H\cos\omega t$ along $x$ and the weak static field $H_z$ along $z$, and drifts with the average velocity $\langle\dot X\rangle$.}
\label{fig:schematic}
\end{revblock}
\end{figure}

We consider the effect of an external magnetic field that oscillates along the $x$ direction, supplemented by a small but finite static field along $z$ \new{(Fig.~\ref{fig:schematic})}: 
\[
\mathbf{H}(t) = (H\cos{\omega t},\,0,\,H_z).
\] 
The weak $z$ field will play a crucial role in symmetry breaking, as explained below. In this section we focus on the ferrimagnetic regime away from spin angular momentum compensation, so that $\delta s$ is finite and the conventional gyrotropic coupling between $X$ and $\Phi$ dominates the velocity-dependent dynamics. In this regime, the field-induced inertial terms discussed in Sec.~\ref{SecII}, including the Euler-, Coriolis-, and centrifugal-like contributions, are smaller by the exchange parameter $M_sH/A_0$ and will be neglected to leading order. The equations of motion for the collective coordinates $X(t)$ and $\Phi(t)$, derived from Eq.~\eqref{ForceEqoM}, are
\begin{equation}\label{EqX}
    \rho\ddot{X}+\delta s \Delta \dot{\Phi}+\alpha_s\dot{X} =M_{net}\Delta H_z,
\end{equation}
\begin{equation}\label{EqPhi}
\begin{aligned}
    \rho\Delta\ddot{\Phi}-\delta s  \dot{X}+\alpha_s\Delta\dot{\Phi} = &-\frac{\pi}{2}M_{net}\Delta H \cos{\omega t}\sin{\Phi}\\
    &-\frac{\pi}{4}D\sin{\Phi}.
\end{aligned}
\end{equation}
where $\alpha_s=2\alpha s$. The first equation shows that the static out-of-plane field gives the usual Zeeman pressure on the wall, while changes in the azimuthal angle feed back into translation through the gyrotropic term proportional to $\delta s$. The second equation shows that the oscillating in-plane field and DMI control the dynamics of the internal angle $\Phi$.

In the steady, non-precessing regime ($\dot{\Phi}=0$) \cite{Oh2017}, Eq.~\eqref{EqX} gives a constant wall velocity
\begin{equation}
    \dot{X}=\frac{M_{net}\Delta H_z}{\alpha_s}.
\end{equation}
Equation~\eqref{EqPhi}, however, admits such non-precessing solutions only when the static drive is below the instantaneous Walker threshold \cite{Walk74,Mougin2007}. Substituting Eq.~\eqref{EqX} into Eq.~\eqref{EqPhi}, we find
\begin{equation}\label{EqWB}
    H^{WB}(t)=\left|\frac{\pi}{4}\frac{\alpha_s}{\delta s}\Big(\frac{D}{M_{net}\Delta}+2H\cos{\omega t}\Big)\right|.
\end{equation}
Unlike the static case, this threshold is time dependent because the in-plane field periodically changes the effective restoring torque on $\Phi$. Thus, even when the static field $H_z$ is small, the system can enter the precessional regime during portions of the cycle for which $H_z>H^{WB}(t)$. \new{This requires the in-plane field to bring the instantaneous threshold to zero during part of the cycle. According to Eq.~\eqref{EqWB}, $H^{WB}(t)$ vanishes when $H\cos\omega t=-H_D$, where $H_D=D/(2M_{net}\Delta)$ is the field at which the torque of the in-plane field on $\Phi$ balances the DMI torque. This is possible only if $H>H_D$.}

For \new{$H>H_D$ and} small $H_z$, the azimuthal angle $\Phi$ switches periodically between $0$ and $\pi$, closely following the oscillating in-plane field with a lag set by damping and DMI. During each switching event, $\dot{\Phi}$ can take either sign, corresponding to clockwise or counterclockwise rotation viewed from the $+z$ axis. When $H_z=0$, the two switching directions are symmetry related, so successive switching events do not produce a preferred angular motion and the long-time average vanishes, $\langle\dot{\Phi}\rangle=0$. A finite $H_z$ breaks this symmetry by driving the wall translationally; through the gyrotropic coupling in Eq.~\eqref{EqPhi}, this translational motion biases the sign of $\dot{\Phi}$. As a result, the switching events acquire a preferred sense of rotation, and $\langle\dot{\Phi}\rangle$ becomes nonzero with a sign selected by $H_z$.

The coupling between azimuthal and translational dynamics, represented by the nonzero $\delta s$ terms in Eqs.~\eqref{EqX}--\eqref{EqPhi}, means that azimuthal switching induces an additional translational contribution. Averaging Eq.~\eqref{EqX} over times much longer than $2\pi/\omega$ and setting $\langle\ddot{X}\rangle=0$, we obtain
\begin{equation}\label{EqAvX}
    \langle\dot{X}\rangle=\frac{M_{net}\Delta H_z}{\alpha_s}-\frac{\delta s \Delta}{\alpha_s}\langle\dot{\Phi}\rangle.
\end{equation}
The first term is the usual field-driven velocity from the static out-of-plane field, while the second arises from the winding of the internal domain-wall angle. This second contribution vanishes at spin angular momentum compensation, $\delta s=0$, within the leading-order approximation used here. 

This behavior is closely related to the phase-locking regime discussed for rotating or circularly polarized magnetic fields \cite{Jin2021,Pan2018}. In those cases, the applied field direction rotates continuously and can lock the domain-wall angle to the drive. Here the applied in-plane field is linearly oscillating rather than circularly polarized, so the field direction switches between $+\hat{\mathbf x}$ and $-\hat{\mathbf x}$. In the locked switching regime, $\Phi(t)$ follows these reversals with a fixed sense selected by the weak static $H_z$, advancing by approximately one full $2\pi$ cycle per drive period. Thus $\langle\dot{\Phi}\rangle\approx\sigma\omega$, where $\sigma=\mathrm{sgn}\,H_z$ gives the sense of rotation.
Equation~\eqref{EqAvX} then gives
\begin{equation}\label{EqAvgV}
    \langle\dot{X}\rangle=\frac{M_{net}\Delta H_z}{\alpha_s}-\sigma\frac{\delta s \Delta}{\alpha_s}\omega.
\end{equation}
\new{We refer to the second term as the frequency-driven drift, $\langle\dot X\rangle_\omega=-\sigma\delta s\Delta\omega/\alpha_s$.} \new{It} is linear in the oscillation frequency rather than the field amplitude\new{, and its direction is reversed by reversing $H_z$}. \new{This drift} is a rectified contribution\new{. The} linearly oscillating in-plane field repeatedly switches the internal angle, while the weak static $H_z$ selects a preferred switching direction. \new{Although the wall angle is driven differently, the frequency-driven drift has the same form as the locked velocity of a ferrimagnetic wall under a rotating field \cite{Jin2021}. In the ferromagnetic limit, $\delta s = 2s$, it becomes $\Delta\omega/\alpha$. This has the same form as the velocity of a ferromagnetic wall under a linearly oscillating field in a regime where the static field exceeds the Walker field at all times and the oscillating field locks the resulting precession \cite{Kim2020PRB}, whereas here the static field exceeds the instantaneous threshold $H^{WB}(t)$ only during part of each cycle.}

\new{At high frequencies, the wall angle can no longer follow the drive. To find where this happens, we neglect the inertial terms $\rho\ddot X$ and $\rho\Delta\ddot\Phi$ in Eqs.~\eqref{EqX}--\eqref{EqPhi} (see Appendix~\ref{App:Cutoff}), and eliminate $\dot X$ from Eq.~\eqref{EqPhi} using Eq.~\eqref{EqX}. This gives
\begin{equation}\label{EqPhiRed}
\dot\Phi=\omega_z-\omega_H\left(\cos\omega t+c\right)\sin\Phi ,
\end{equation}
where the bias frequency $\omega_z$ and the switching frequency $\omega_H$ are
\begin{equation}
\omega_z=\frac{\delta s M_{net}H_z}{\alpha_s^2+\delta s^2},\qquad
\omega_H=\frac{\pi}{2}\frac{\alpha_s M_{net} H}{\alpha_s^2+\delta s^2},
\end{equation}
and $c=H_D/H$. The restoring torque on $\Phi$, proportional to $\cos\omega t+c$, changes sign during the cycle only if $H>H_D$, as found above from Eq.~\eqref{EqWB}. Below this threshold, $\Phi$ stays near its equilibrium orientation and the wall does not switch. Above it, the stable orientation of the wall angle alternates during each period. It is $\Phi=\pi$ while $\cos\omega t<-c$ and $\Phi=0$ for the rest of the period. The wall angle turns by $\pi$ each time the stable orientation changes. Each turn can go either way. At the start of a turn, $\Phi$ sits at an orientation that has just become unstable. The restoring torque vanishes there, so even the weak bias $\omega_z$ selects the sense of the turn. When both turns in a period follow the bias, $\Phi$ advances by $2\pi$ per period and the wall is locked to the drive. This requires $\Phi$ to settle at each stable orientation before it becomes unstable. At high frequencies, the orientation $\Phi=\pi$ is not stable long enough for the wall angle to reach it. The next turn then starts slightly on the side of $\Phi=0$. The wall stays locked only as long as the bias can overcome this offset. As shown in Appendix~\ref{App:Cutoff}, this condition gives the cutoff frequency $\omega_c$ as the solution of
\begin{equation}\label{EqCutoff}
\frac{2\omega_H}{\omega_c}f(c)=\ln\!\left[\frac{2}{\pi}\,\frac{\omega_c\,\omega_H\sqrt{1-c^2}}{\omega_z^2}\right],
\end{equation}
with $f(c)=\sqrt{1-c^2}-c\arccos c$. Equation~\eqref{EqCutoff} contains no fitting parameters. It assumes that the torque passes through zero at two well-separated instants in each period. As $H$ approaches $H_D$ and $c$ approaches 1, the reversed part of the period shrinks and these two instants merge. Equation~\eqref{EqCutoff} then no longer applies. The cutoff increases with $H$ through $\omega_H$ and $c$ and with $H_z$ through $\omega_z$. Above $\omega_c$, the switching becomes incomplete or irregular, $|\langle\dot{\Phi}\rangle|$ drops below $\omega$, and the frequency-driven drift is suppressed.}

\subsection{Numerical Analysis}

\new{To confirm our analytical results, we performed micromagnetic simulations of a synthetic ferrimagnet using \textsc{mumax3} \cite{Vansteenkiste2014,Mulkers2017}. The sample is a strip of $1024\times16$~nm$^2$ made of two 1-nm-thick ferromagnetic layers, one for each sublattice, coupled antiferromagnetically through an interlayer exchange equal in magnitude to the intralayer exchange. The strip is discretized into $1\times1\times1$~nm$^3$ cells, with open boundary conditions. The two layers have exchange stiffnesses $A_1=A_2=10$~pJ/m, perpendicular uniaxial anisotropies $K_1=K_2=650$~kJ/m$^3$, interfacial DMI constants $D_1=D_2=0.03$~mJ/m$^2$, saturation magnetizations $M_1=900$~kA/m and $M_2=1010$~kA/m, gyromagnetic ratios $\gamma_1=\gamma_2=1.7595\times10^{11}$~rad\,s$^{-1}$T$^{-1}$, and Gilbert damping constants $\alpha_1=\alpha_2=0.02$. These give a net spin density $\delta s=-0.625$~$\mu$J$\cdot$s/m$^3$. As in the analytical model, we neglect the dipolar interaction, which is weak in ferrimagnets because of their small net magnetization \cite{Kim2022NatMater}. The in-plane field has amplitude $H=200$~mT, and the static out-of-plane field is $H_z=5$~mT. The initial state is a domain wall relaxed in zero field, placed so that it stays at least 50~nm from the ends of the strip during the run. The LLG equation is integrated with the adaptive Dormand--Prince solver of \textsc{mumax3}. Each run lasts between 10 and 83~ns, depending on the drive frequency, so that it covers at least one full drive period after an initial transient of at least 1~ns. The wall position is obtained from the layer-averaged $m_z$, and the average velocity is the wall displacement over an integer number of drive periods divided by their duration.}

\begin{figure}
\begin{revblock}
\centering
\includegraphics[width=0.58\columnwidth]{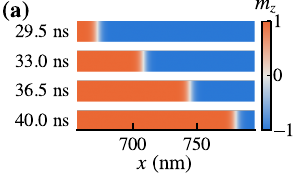}\hfill
\includegraphics[width=0.41\columnwidth]{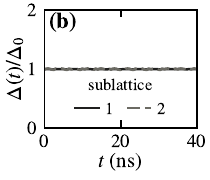}
\caption{Wall rigidity at $\omega=18\times10^8$~rad/s, just below the cutoff, for $H=200$~mT, $H_z=5$~mT and $\delta s=-0.625$~$\mu$J$\cdot$s/m$^3$. (a) $m_z$ of sublattice 1 at four times separated by one drive period, at the same phase of the drive. The wall advances by about 36~nm per period, a net velocity of 10.28~m/s. This includes the static drift of $-9.85$~m/s produced by $H_z$. The frequency-driven part, 20.13~m/s, is the quantity plotted in Fig.~\ref{fig:DW_sims}(a). (b) Wall width of both sublattices, normalized by $\Delta_0=\Delta=\sqrt{A/K_z}=3.92$~nm, over the whole run.}
\label{fig:rigidity}
\end{revblock}
\end{figure}

\begin{figure*}
\begin{revblock}
\centering
\includegraphics[width=0.5\textwidth]{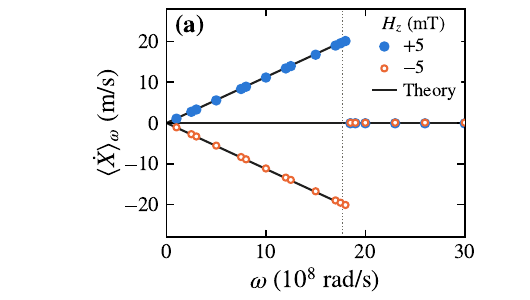}%
\includegraphics[width=0.5\textwidth]{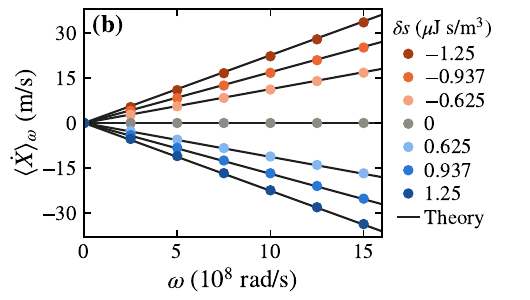}\\
\includegraphics[width=0.5\textwidth]{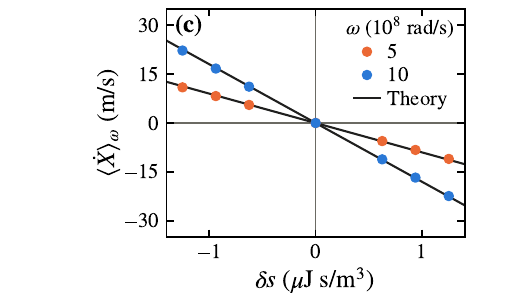}%
\includegraphics[width=0.5\textwidth]{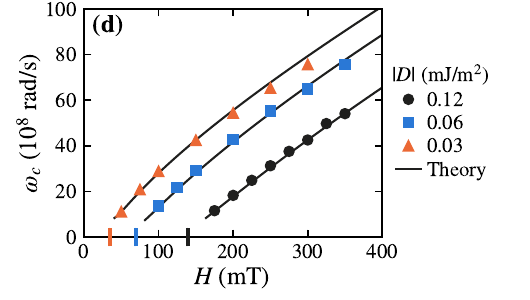}
\caption{Symbols are \textsc{mumax3} simulations and lines are theory. (a)--(c) Frequency-driven drift $\langle\dot X\rangle_\omega$, the average wall velocity minus the drift due to $H_z$, for $H=200$~mT. Lines are Eq.~\eqref{EqAvgV}. (a) Versus $\omega$ at $\delta s=-0.625$~$\mu$J$\cdot$s/m$^3$ for $H_z=\pm5$~mT. The dotted line is the cutoff predicted by Eq.~\eqref{EqCutoff}. (b) Versus $\omega$ for several $\delta s$, with $M_1+M_2=1910$~kA/m. (c) Versus $\delta s$ at two drive frequencies. (d) Cutoff frequency versus $H$ for three DMI strengths, $|D|=2(D_1+D_2)$, at $\delta s=-0.625$~$\mu$J$\cdot$s/m$^3$ and $H_z=5$~mT. Lines are Eq.~\eqref{EqCutoff}, drawn for $H_D/H\le0.85$, and the ticks on the horizontal axis mark $H_D$.}
\label{fig:DW_sims}
\end{revblock}
\end{figure*}

\new{Because the simulated system is a synthetic ferrimagnet, its parameters must be renormalized before they can be compared with our continuum model. As shown in Appendix~\ref{App:SynthFiM}, the values above correspond to $A=2(A_1+A_2)=40$~pJ/m, $K_z=2(K_1+K_2)=2.6$~MJ/m$^3$ and $|D|=2(D_1+D_2)=0.12$~mJ/m$^2$ in Eq.~\eqref{Eq:EnergyDensity}, with $M_{net}=M_1-M_2=-110$~kA/m, which gives a wall width $\Delta=\sqrt{A/K_z}=3.92$~nm. These are the values used in the theory curves below. Where other values of the DMI, the sublattice magnetizations or the applied fields are used, they are stated with the corresponding figures.}

\new{Before comparing the simulated velocities with Eq.~\eqref{EqAvgV}, we check the central assumption of the analytical model, that the wall moves as a rigid object of nearly constant width. Figure~\ref{fig:rigidity} shows a run at $\omega=18\times10^8$~rad/s, the highest simulated frequency below the cutoff for $H=200$~mT. Figure~\ref{fig:rigidity}(a) shows $m_z$ of sublattice 1 at four times separated by one drive period, all at the same phase of the drive. The wall keeps its shape and remains straight across the strip while it advances by about 36~nm per period. Figure~\ref{fig:rigidity}(b) shows that the width of each sublattice wall stays within 2.5\% of the model width $\Delta=3.92$~nm throughout the run, with only small oscillations. We also checked the shape of the wall. At all times, the $m_z$ profile of each sublattice agrees with $\pm\tanh[(x-X)/\Delta]$ from Eq.~\eqref{Eq:DWansatz}, with a root-mean-square deviation below $10^{-3}$. The walls of the two sublattices coincide to within 0.02~nm, and their magnetizations stay antiparallel to within $1^\circ$. The same holds at $\omega=5$, 17 and $20\times10^8$~rad/s, the last just above the cutoff. The rigid-wall description therefore holds, including near the cutoff.}

\new{Figures~\ref{fig:DW_sims}(a)--\ref{fig:DW_sims}(c) show the frequency-driven drift $\langle\dot X\rangle_\omega$. We obtain it by subtracting the drift due to $H_z$ from the simulated average velocity. In Fig.~\ref{fig:DW_sims}(a), $\delta s=-0.625$~$\mu$J$\cdot$s/m$^3$ and the drive frequency extends above the cutoff, for both signs of the static field. Below the cutoff, $\langle\dot X\rangle_\omega$ is linear in the drive frequency and follows Eq.~\eqref{EqAvgV} without any fitting parameter. Reversing $H_z$ reverses $\langle\dot X\rangle_\omega$, while its magnitude and the cutoff remain the same, as expected from $\sigma=\mathrm{sgn}\,H_z$. This confirms the main analytical prediction. In the locked switching regime, the linearly oscillating field drives the wall through the frequency of the repeated switching events rather than through its amplitude, and the static field sets the direction of motion. Above the cutoff predicted by Eq.~\eqref{EqCutoff}, shown by the dotted line, $\langle\dot X\rangle_\omega$ drops to zero.}

\new{To test the dependence on the net spin density, we varied $M_1$ and $M_2$ while keeping $M_1+M_2=1910$~kA/m fixed, so that $\alpha_s$ stays the same and $\delta s$ ranges from $-1.25$ to $1.25$~$\mu$J$\cdot$s/m$^3$. As shown in Fig.~\ref{fig:DW_sims}(b), the slope of $\langle\dot X\rangle_\omega$ versus $\omega$ is proportional to $\delta s$ and changes sign with it, as predicted by Eq.~\eqref{EqAvgV}. Figure~\ref{fig:DW_sims}(c) shows the drift as a function of $\delta s$ at two drive frequencies. The drift is linear in $\delta s$ and vanishes at spin angular momentum compensation, $\delta s=0$, which confirms that the frequency-driven drift is gyrotropic in origin. Although Sec.~\ref{SecII} shows that the field-induced Coriolis-like terms can couple $X$ and $\Phi$ even at compensation, for the spatially uniform field considered here they and the Euler-like terms give only oscillatory corrections over a closed switching cycle, so the absence of drift at $\delta s=0$ is expected.}

\new{We also determined the cutoff frequency as a function of the in-plane field amplitude for three DMI strengths, $D_1=D_2=0.03$, $0.015$ and $0.0075$~mJ/m$^2$, which correspond to $|D|=0.12$, $0.06$ and $0.03$~mJ/m$^2$ and to threshold fields $H_D=139$, $70$ and $35$~mT. Figure~\ref{fig:DW_sims}(d) shows the result. For each DMI strength, the cutoff drops toward zero as $H$ approaches $H_D$ and grows approximately linearly with $H$ above it. Equation~\eqref{EqCutoff} agrees well with the simulated cutoffs for all three DMI strengths, without any fitting parameter. The simulations thus confirm the two ingredients of the rectified drift. The in-plane field must exceed $H_D$ to switch $\Phi$, and the drive frequency must remain below $\omega_c$ so that the switching stays locked to the oscillation period. Since $H_D=D/(2M_{net}\Delta)$, the ferrimagnetic counterpart of the DMI field of ferromagnetic walls \cite{Thiaville2012}, a weaker DMI or a wider wall lowers the field required for the drift. For $|D|=0.03$~mJ/m$^2$, the wall is already driven at $H=50$~mT, with a cutoff of $11\times10^8$~rad/s. In a single GdFeCo layer, for example, the DMI is about $0.01$~mJ/m$^2$ and the walls are about 20~nm wide \cite{Krishnia2021}, so $H_D$ is of the order of 10~mT, which a single on-chip stripline can supply \cite{Bisig2015}.}

\section{Summary} \label{SecIV}

We have derived the collective-coordinate equations of motion for ferrimagnetic domain walls in the presence of time-dependent magnetic fields, retaining the field-dependent contribution to the kinetic energy. This formulation naturally produces Euler, Coriolis, and centrifugal-like terms in the domain-wall equations of motion. The Euler-like term is associated with the time derivative of the applied field, while the Coriolis-like term appears as a field-induced velocity-dependent coupling between the domain-wall position and azimuthal angle. Although these field-induced inertial terms are small in the exchange limit, they become conceptually important near magnetic compensation, where conventional ferrimagnetic terms can be suppressed.

We then studied the dynamics of a ferrimagnetic domain wall driven by a linearly oscillating in-plane magnetic field supplemented by a weak static out-of-plane field. Away from spin angular momentum compensation, the leading dynamics are governed by the conventional gyrotropic coupling proportional to the uncompensated spin density $\delta s$, while the field-induced inertial terms give only higher-order corrections. The oscillating in-plane field periodically modifies the effective Walker threshold and, \new{once its amplitude exceeds the threshold $H_D$ set by the DMI,} drives repeated switching of the domain-wall azimuthal angle. The weak static out-of-plane field breaks the symmetry between clockwise and counterclockwise switching, producing a preferred angular motion. Through the gyrotropic coupling between $X$ and $\Phi$, this rectified azimuthal switching generates a finite average domain-wall velocity.

\new{Our analytical results show that, in the locked switching regime, the frequency-driven drift is proportional to the oscillation frequency rather than to the field amplitude, and its direction is reversed by reversing the static field. Above a cutoff frequency, for which we derived an expression without fitting parameters, the wall angle can no longer follow the drive and the drift is suppressed. Micromagnetic simulations confirm these predictions. They show the linear dependence of the drift on the drive frequency, its reversal with the static field and with the net spin density, and cutoff frequencies that agree well with the analytical expression for three DMI strengths.} At spin angular momentum compensation, $\delta s=0$, the \new{frequency-driven} drift vanishes in the leading-order mechanism considered here, consistent with the absence of conventional gyrotropic coupling.

Several future directions follow from this work. First, the field-induced Coriolis-like tensor identified in Sec.~\ref{SecII} may provide an additional mechanism for coupling translational and azimuthal dynamics near compensation, where the conventional gyrotropic tensor vanishes. In the present linearly oscillating and spatially uniform field geometry, this coupling does not produce a dc average velocity over a closed switching cycle, but more general driving protocols could reveal its effects. Examples include spatially nonuniform fields, chiral or twisted domain-wall textures, and rotating or elliptically polarized fields. Second, it would be useful to study the near-compensation regime quantitatively, where the conventional gyrotropic coupling and the field-induced inertial terms may compete. \new{Finally, the effect of pinning and finite sample size on the threshold amplitude and the cutoff frequency remains to be studied.}

\begin{acknowledgments}
We thank Mohit Randeria and Denis Pelekhov for very useful discussions. We acknowledge support from the NSF Materials Research Science and Engineering Center Grant No.~DMR-2011876. M.M. was also supported by The Ohio State University's Distinguished University Fellowship.
\end{acknowledgments}

\appendix

\section{Ferrimagnetic Lagrangian}\label{App:FiMLagrangian}

In this appendix, we derive the effective ferrimagnetic Lagrangian expressed in Eq.\eqref{Eq:FiMLagrangian_mn} in the main text starting from the two-sublattice LLG equations. In particular, we show how the spin Berry-phase terms of the two antiferromagnetically coupled sublattices reduce, in the exchange-dominated limit, to an antiferromagnetic-like kinetic term and a ferromagnetic-like Berry-phase term proportional to the net spin density $\delta s$. We also derive the corresponding Rayleigh dissipation function used to obtain the damping terms in the collective-coordinate equations.

Starting from Eq.~\eqref{Eq.LLG}, we write $\mathbf{M}_i=M_{is}\mathbf{m}_i$, where $M_{is}$ is assumed to be constant and $\mathbf{m}_i$ is a unit vector. Introducing the spin density $s_i=M_{is}/\gamma_i$, the LLG equation becomes
\begin{equation}\label{App:LLGang}
    s_i\frac{\partial \mathbf{m}_i}{\partial t}
    =
    \mathbf{m}_i\times\frac{\delta U}{\delta \mathbf{m}_i}
    +\alpha_i s_i\,\mathbf{m}_i\times
    \frac{\partial \mathbf{m}_i}{\partial t}.
\end{equation}
This equation follows from the spin Berry-phase Lagrangian density \cite{Auerbach1994,Tatara2008}
\begin{equation}
    \mathcal{L}_i
    =
    s_i\boldsymbol{\mathcal A}_i(\mathbf{m}_i)\cdot
    \partial_t\mathbf{m}_i
    -\mathcal{U},
\end{equation}
where $\boldsymbol{\nabla}_{m_i}\times\boldsymbol{\mathcal A}_i(\mathbf{m}_i)=\mathbf{m}_i$ is the Berry connection on the unit sphere \cite{Tserk2017,Tatara2008} and $U=\int dV\,\mathcal{U}$ is the total magnetic energy. Gilbert damping can be included through the Rayleigh dissipation function \cite{Tatara2008}
\begin{equation}
    R_i=\frac{1}{2}\alpha_i s_i
    \left|\partial_t\mathbf{m}_i\right|^2 .
\end{equation}
The corresponding Euler-Lagrange equation with dissipation is
\begin{equation}\label{App:EqELR}
    \frac{\partial \mathcal{L}}{\partial \mathbf{m}_i}
    -\frac{\partial}{\partial t}
    \frac{\partial \mathcal{L}}{\partial(\partial_t\mathbf{m}_i)}
    -\boldsymbol{\nabla}\cdot
    \frac{\partial \mathcal{L}}{\partial(\boldsymbol{\nabla}\mathbf{m}_i)}
    =
    \frac{\partial R}{\partial(\partial_t\mathbf{m}_i)} .
\end{equation}

For a ferrimagnet with two antiferromagnetically coupled sublattices, the Lagrangian density is obtained by summing the Berry-phase terms:
\begin{equation}
    \mathcal{L}
    =
    s_1\boldsymbol{\mathcal A}_1(\mathbf{m}_1)\cdot\partial_t\mathbf{m}_1
    +
    s_2\boldsymbol{\mathcal A}_2(\mathbf{m}_2)\cdot\partial_t\mathbf{m}_2
    -\mathcal{U}.
\end{equation}
We introduce the net and staggered fields $\mathbf{m}=\mathbf{m}_1+\mathbf{m}_2$, $\mathbf{n}=\frac{\mathbf{m}_1-\mathbf{m}_2}{2}$, so that $\mathbf{m}_1=\mathbf{m}/2+\mathbf{n}$ and $\mathbf{m}_2=\mathbf{m}/2-\mathbf{n}$. In the exchange-dominated regime, $|\mathbf{m}|\ll1$ and $|\mathbf{n}|\simeq1$. Expanding the two Berry-phase terms to leading order in $\mathbf{m}$ and choosing opposite Dirac-string gauges for the nearly antiparallel sublattices gives
\begin{equation}
    \mathcal{L}
    =
    s\,\partial_t\mathbf{n}\cdot(\mathbf{n}\times\mathbf{m})
    +
    \delta s\,\boldsymbol{\mathcal A}(\mathbf{n})\cdot\partial_t\mathbf{n}
    -\mathcal{U},
\end{equation}
where $s=(s_1+s_2)/2$ and $\delta s=s_1-s_2$ . The first term describes antiferromagnetic-like inertial dynamics, while the second is the Berry-phase contribution from the uncompensated spin density. At spin angular momentum compensation, $\delta s=0$, the ferromagnetic-like Berry-phase term vanishes.

The Rayleigh dissipation function is expanded in the same variables:
\begin{equation}
\begin{aligned}
    R
    &=
    \frac{1}{2}s_1\alpha_1
    \left|\partial_t\mathbf{m}_1\right|^2
    +
    \frac{1}{2}s_2\alpha_2
    \left|\partial_t\mathbf{m}_2\right|^2  \\
    &=
    \frac{1}{2}\alpha_s
    \left|\partial_t\mathbf{n}\right|^2
    +\mathcal{O}(\partial_t\mathbf{n}\cdot\partial_t\mathbf{m}),
\end{aligned}
\end{equation}
where $\alpha_s=s_1\alpha_1+s_2\alpha_2$. To leading order in the exchange approximation, $R\simeq \frac{1}{2}\alpha_s \left|\partial_t\mathbf{n}\right|^2$. For equal damping constants, $\alpha_1=\alpha_2=\alpha$, one has $\alpha_s=2\alpha s$. 

\section{Ferrimagnetic Lagrangian with External Field}\label{App:FiMLagrangianField}

In this appendix, we derive the effective ferrimagnetic Lagrangian in the presence of an external magnetic field. Starting from the ferrimagnetic Lagrangian obtained in Appendix~\ref{App:FiMLagrangian}, we include the Zeeman coupling to both the small magnetization $\mathbf{m}$ and the staggered order parameter $\mathbf{n}$. We then integrate out $\mathbf{m}$ in the exchange-dominated limit, which gives the field-dependent kinetic term used in the main text.

The magnetic energy density is
\begin{equation}
\begin{aligned}
    \mathcal{U}
    &=
    \frac{A_0}{2}\mathbf{m}^2
    +\frac{A}{2}(\partial_x \mathbf{n})^2
    -\frac{K_z}{2}n_z^2
    \\&-\frac{D}{2}\hat{\mathbf y}\cdot(\mathbf{n}\times\partial_x \mathbf{n})
    -\mathbf{H}\cdot\mathbf{M},
\end{aligned}
\end{equation}
where the last term is the Zeeman coupling. We use $\mathbf{M}=s\gamma\,\mathbf{m}+M_{\mathrm{net}}\mathbf{n}$, where $M_{\mathrm{net}}=M_{1s}-M_{2s}$ is the net magnetization, $s=(s_1+s_2)/2$, and $\gamma=(\gamma_1s_1+\gamma_2s_2)/(s_1+s_2)$. Substituting the Zeeman coupling into the ferrimagnetic Lagrangian gives
\begin{equation}\label{App:EqLagField}
\begin{aligned}
    \mathcal{L}_{\mathrm{FiM}}
    &=
    s\,\partial_t\mathbf{n}\cdot(\mathbf{n}\times\mathbf{m})
    +\delta s\,\boldsymbol{\mathcal A}(\mathbf{n})\cdot\partial_t\mathbf{n}
    -\frac{A_0}{2}\mathbf{m}^2
    \\
    &\quad
    -\frac{A}{2}(\partial_x \mathbf{n})^2
    +\frac{K_z}{2}n_z^2
    +\frac{D}{2}\hat{\mathbf y}\cdot(\mathbf{n}\times\partial_x \mathbf{n})
    \\
    &\quad
    +s\gamma\,\mathbf{m}\cdot\mathbf{H}
    +M_{\mathrm{net}}\mathbf{n}\cdot\mathbf{H}.
\end{aligned}
\end{equation}
This form is the exchange-dominated ferrimagnetic Lagrangian supplemented by the Zeeman coupling to the external magnetic field \cite{Tveten2016,Tserk2017}.

The small magnetization $\mathbf{m}$ is not an independent slow degree of freedom in the exchange approximation. It satisfies the constraints $\mathbf{n}^2=1$ and $\mathbf{m}\cdot\mathbf{n}=0$. Varying Eq.~\eqref{App:EqLagField} with respect to $\mathbf{m}$ and projecting perpendicular to $\mathbf{n}$ gives
\begin{equation}\label{App:mField}
    \mathbf{m}
    =
    \frac{s}{A_0}\left(\partial_t\mathbf{n}\times\mathbf{n}\right)
    +
    \frac{s\gamma}{A_0}
    \left(\mathbf{n}\times\mathbf{H}\right)\times\mathbf{n}.
\end{equation}
Thus, $\mathbf{m}$ contains a dynamical contribution generated by the motion of $\mathbf{n}$ and a transverse magnetization induced by the applied field.

Substituting Eq.~\eqref{App:mField} back into Eq.~\eqref{App:EqLagField}, we obtain
\begin{equation}\label{App:FiMLagrangianFieldFinal}
    \mathcal{L}_{\mathrm{FiM}}
    =
    \frac{\rho}{2}
    \left(
    \partial_t\mathbf{n}
    -\gamma\mathbf{H}\times\mathbf{n}
    \right)^2
    +
    \delta s\,\boldsymbol{\mathcal A}(\mathbf{n})\cdot\partial_t\mathbf{n}
    -\mathcal{U}(\mathbf{n}),
\end{equation}
where $\rho=s^2/A_0$, and
\[
    \mathcal{U}(\mathbf{n})
    =
    \frac{A}{2}(\partial_x \mathbf{n})^2
    -\frac{K_z}{2}n_z^2
    -\frac{D}{2}\hat{\mathbf y}\cdot(\mathbf{n}\times\partial_x \mathbf{n})
    -M_{\mathrm{net}}\mathbf{n}\cdot\mathbf{H}.
\]
Equation~\eqref{App:FiMLagrangianFieldFinal} shows that the external magnetic field enters the kinetic energy through the covariant time derivative $\partial_t\mathbf{n}\rightarrow\partial_t\mathbf{n}-\gamma\mathbf{H}\times\mathbf{n}$. This field-dependent kinetic term is the origin of the Euler-, Coriolis-, and centrifugal-like contributions discussed in Sec.~\ref{SecII}.

\section{Synthetic Ferrimagnet}\label{App:SynthFiM}

In this appendix, we show that the synthetic ferrimagnet used in the micromagnetic simulations reduces to the continuum ferrimagnetic model used in the main text. The synthetic ferrimagnet consists of two ferromagnetic layers with different magnetizations, coupled antiferromagnetically across the layers. In the exchange-dominated limit, the only effect of this synthetic realization is a renormalization of the exchange, anisotropy and DMI parameters appearing in the effective ferrimagnetic Lagrangian.

We denote the magnetization directions in the two layers by $\mathbf{m}_1$ and $\mathbf{m}_2$. Each layer has an intralayer ferromagnetic exchange stiffness $A_i$ with $i=1,2$, while the two layers are coupled by an antiferromagnetic interlayer exchange $A_{\mathrm{AF}}>0$. In the convention of \textsc{mumax3} \cite{Vansteenkiste2014}, the continuum exchange-energy density is
\begin{equation}
     \mathcal{H}_{\mathrm{ex}}
     =
     A_{\mathrm{AF}}\mathbf{m}_1\cdot\mathbf{m}_2
     +A_1(\partial_x\mathbf{m}_1)^2
     +A_2(\partial_x\mathbf{m}_2)^2 .
\end{equation}
This form follows from a lattice model with ferromagnetic intralayer exchange and antiferromagnetic interlayer exchange after taking the continuum limit.

We now introduce the net and staggered fields used throughout the main text,
$\mathbf{m}=\mathbf{m}_1+\mathbf{m}_2$ and
$\mathbf{n}=(\mathbf{m}_1-\mathbf{m}_2)/2$, so that
$\mathbf{m}_1=\mathbf{m}/2+\mathbf{n}$ and
$\mathbf{m}_2=\mathbf{m}/2-\mathbf{n}$. The constraints are $\mathbf{n}^2+\mathbf{m}^2/4=1$ and $\mathbf{m}\cdot\mathbf{n}=0$. Therefore, to leading order in $|\mathbf{m}|\ll1$, the interlayer exchange term becomes
$A_{\mathrm{AF}}\mathbf{m}_1\cdot\mathbf{m}_2
=
A_{\mathrm{AF}}(\mathbf{m}^2/2-1)$, up to an irrelevant constant. The gradient terms become
\begin{equation}
\begin{aligned}
    A_1(\partial_x\mathbf{m}_1)^2
   & +A_2(\partial_x\mathbf{m}_2)^2
    =
    (A_1+A_2)(\partial_x\mathbf{n})^2\\
   & +\frac{A_1+A_2}{4}(\partial_x\mathbf{m})^2
    +(A_1-A_2)\,\partial_x\mathbf{m}\cdot\partial_x\mathbf{n}.
\end{aligned}
\end{equation}
In the exchange approximation, $\mathbf{m}$ is small, so terms involving spatial gradients of $\mathbf{m}$ are higher order and can be neglected \cite{Tveten2016,Nakane2021}. The exchange-energy density then reduces to
\begin{equation}
    \mathcal{H}_{\mathrm{ex}}
    \simeq
    \frac{A_{\mathrm{AF}}}{2}\mathbf{m}^2
    +
    (A_1+A_2)(\partial_x\mathbf{n})^2 .
\end{equation}
Comparing with Eq.~\eqref{Eq:EnergyDensity}, the synthetic ferrimagnet has the same continuum form as the ferrimagnetic model used in the main text, with $A_0=A_{\mathrm{AF}}$ and $A=2(A_1+A_2)$.

The anisotropy and interfacial DMI energy densities of layer $i$ are $-K_i m_{i,z}^2$ and $D_i(m_{i,z}\partial_x m_{i,x}-m_{i,x}\partial_x m_{i,z})$ \cite{Mulkers2017}. Both are quadratic in $\mathbf{m}_i$, so setting $\mathbf{m}_1\simeq\mathbf{n}$ and $\mathbf{m}_2\simeq-\mathbf{n}$ gives $-(K_1+K_2)n_z^2$ and $(D_1+D_2)(n_z\partial_x n_x-n_x\partial_x n_z)$. Comparing with Eq.~\eqref{Eq:EnergyDensity} yields $K_z=2(K_1+K_2)$ and $D=-2(D_1+D_2)$. For the parameters of Sec.~\ref{SecIII}, $A=40$~pJ/m, $K_z=2.6$~MJ/m$^3$ and $D=-0.12$~mJ/m$^2$, so that $\Delta=\sqrt{A/K_z}=3.92$~nm, close to the simulated static width of 3.88~nm. The sign of $D$ only sets the chirality of the wall, so we quote its magnitude in the main text. The synthetic nature of the simulated system therefore only renormalizes these parameters and does not change the structure of the effective ferrimagnetic dynamics.

\section{Cutoff frequency}\label{App:Cutoff}

In this appendix, we justify Eq.~\eqref{EqPhiRed} and derive Eq.~\eqref{EqCutoff} from it. Equation~\eqref{EqPhiRed} neglects the inertial terms $\rho\ddot X$ and $\rho\Delta\ddot\Phi$ of Eqs.~\eqref{EqX}--\eqref{EqPhi}. With $\rho=s^2/A_0$ and $A_0\sim10^7$~J/m$^3$ for the synthetic ferrimagnet of Sec.~\ref{SecIII}, these terms act on the time scale $\rho/\sqrt{\alpha_s^2+\delta s^2}\sim5\times10^{-12}$~s, an order of magnitude shorter than the switching time $1/\omega_H$ and two orders of magnitude shorter than the drive period. They can therefore be dropped.

We take $\omega_z$, $\omega_H$ and $c$ positive, with $0<c<1$; the other sign combinations follow from the symmetry $\Phi\to-\Phi$ or from a shift of time by half a period. We write $r(t)=\omega_H(\cos\omega t+c)$, so that $\dot\Phi=\omega_z-r(t)\sin\Phi$. Within each period, $r(t)<0$ in the window $t_0<t<t_1$, where $\cos\omega t_0=\cos\omega t_1=-c$ and $\omega(t_1-t_0)=2\arccos c$. Before $t_0$, $\Phi$ stays close to the stable orientation $\Phi=0$. At $t_0$, this orientation becomes unstable, and $\Phi$ moves toward the new stable orientation $\Phi=\pi$. At $t_1$, $\Phi=\pi$ becomes unstable in turn, and $\Phi$ either continues to $2\pi$ or returns to $0$.

Near $t_0$, we expand $r(t)\simeq-\beta(t-t_0)$, with $\beta=\omega\,\omega_H\sqrt{1-c^2}$, and use $\sin\Phi\simeq\Phi$. Equation~\eqref{EqPhiRed} becomes $\dot\Phi=\omega_z+\beta(t-t_0)\Phi$, whose solution for $t-t_0\gg\beta^{-1/2}$ is
\begin{equation}
\Phi(t)\simeq\varepsilon\, e^{\beta(t-t_0)^2/2},\qquad \varepsilon=\omega_z\sqrt{2\pi/\beta}.
\end{equation}
The bias therefore pushes $\Phi$ away from the unstable orientation by an effective amount $\varepsilon$, set by the time $\beta^{-1/2}$ that the angle needs to respond near the crossing.

For $t_0<t<t_1$, the bias is negligible compared with $|r(t)|$, and Eq.~\eqref{EqPhiRed} gives $d\ln\tan(\Phi/2)/dt=-r(t)$. Over the window, $\tan(\Phi/2)$ therefore grows by the factor $e^{G}$, with
\begin{equation}
G=\int_{t_0}^{t_1}|r(t)|\,dt=\frac{2\omega_H}{\omega}f(c),
\end{equation}
with $f(c)$ defined after Eq.~\eqref{EqCutoff}. Starting from $\tan(\Phi/2)\simeq\varepsilon/2$, the angle approaches $\pi$ and arrives at $t_1$ with the deviation $\pi-\Phi\simeq\varepsilon'=(4/\varepsilon)e^{-G}$. This is the offset toward $\Phi=0$ discussed in Sec.~\ref{SecIII}.

Near $t_1$, $r(t)\simeq\beta(t-t_1)$, and with $\Phi=\pi-\eta$ the deviation obeys $\dot\eta=\beta(t-t_1)\eta-\omega_z$. By the same argument as near $t_0$, the bias shifts $\eta$ by $-\varepsilon$, so that after the crossing $\eta$ grows in proportion to $\varepsilon'-\varepsilon$. If $\varepsilon>\varepsilon'$, $\Phi$ passes $\pi$ and completes the turn to $2\pi$, and the wall remains locked. If $\varepsilon<\varepsilon'$, $\Phi$ falls back to $0$ and the winding is lost. The rest of the cycle, from $t_1$ to $t_0+2\pi/\omega$, is longer and has a larger growth factor, so it does not limit the locking. Setting $\varepsilon=\varepsilon'$ gives $G=\ln(4/\varepsilon^2)$, which is Eq.~\eqref{EqCutoff}.

The derivation assumes that the two crossings are well separated, $\beta^{-1/2}\ll t_1-t_0$. This fails as $c$ approaches 1, where $t_1-t_0=2\arccos c/\omega$ vanishes. Away from the threshold, Eq.~\eqref{EqCutoff} agrees well with the cutoff obtained by integrating Eq.~\eqref{EqPhiRed} numerically and with the simulations in Fig.~\ref{fig:DW_sims}(d). This agreement also confirms that the inertial terms are negligible here.

\bibliography{references}

@PREAMBLE{
  "\providecommand{\noopsort}[1]{}"
}

@article{LL35,
  author  = {Landau, L. D. and Lifshitz, E. M.},
  title   = {On the theory of the dispersion of magnetic permeability in ferromagnetic bodies},
  journal = {Phys. Z. Sowjetunion},
  volume  = {8},
  pages   = {153--169},
  year    = {1935}
}

@article{Gilb04,
  author  = {Gilbert, T. L.},
  title   = {A phenomenological theory of damping in ferromagnetic materials},
  journal = {IEEE Transactions on Magnetics},
  volume  = {40},
  pages   = {3443--3449},
  year    = {2004},
  doi     = {10.1109/TMAG.2004.836740}
}

@article{Walk74,
  author  = {Schryer, N. L. and Walker, L. R.},
  title   = {The motion of 180 degree domain walls in uniform dc magnetic fields},
  journal = {Journal of Applied Physics},
  volume  = {45},
  pages   = {5406--5421},
  year    = {1974},
  doi     = {10.1063/1.1663252}
}

@article{Thiele73,
  author  = {Thiele, A. A.},
  title   = {Steady-State Motion of Magnetic Domains},
  journal = {Physical Review Letters},
  volume  = {30},
  pages   = {230--233},
  year    = {1973},
  doi     = {10.1103/PhysRevLett.30.230}
}

@article{Baryakhtar85,
  author  = {Bar'yakhtar, V. G. and Ivanov, B. A. and Chetkin, M. V.},
  title   = {Dynamics of domain walls in weak ferromagnets},
  journal = {Soviet Physics Uspekhi},
  volume  = {28},
  pages   = {563--588},
  year    = {1985},
  doi     = {10.1070/PU1985v028n07ABEH003871}
}

@article{Andreev1980,
  author  = {Andreev, A. F. and Marchenko, V. I.},
  title   = {Symmetry and the macroscopic dynamics of magnetic materials},
  journal = {Soviet Physics Uspekhi},
  volume  = {23},
  pages   = {21--34},
  year    = {1980},
  doi     = {10.1070/PU1980v023n01ABEH004859}
}

@article{Ivanov19,
  author  = {Ivanov, B. A.},
  title   = {Ultrafast spin dynamics and spintronics for ferrimagnets close to the spin compensation point},
  journal = {Low Temperature Physics},
  volume  = {45},
  pages   = {935--963},
  year    = {2019},
  doi     = {10.1063/1.5121265}
}

@article{Tserk2014,
  author  = {Kim, S. K. and Tserkovnyak, Y. and Tchernyshyov, O.},
  title   = {Propulsion of a domain wall in an antiferromagnet by magnons},
  journal = {Physical Review B},
  volume  = {90},
  pages   = {104406},
  year    = {2014},
  doi     = {10.1103/PhysRevB.90.104406}
}

@article{Tserk2017,
  author  = {Kim, S. K. and Lee, K.-J. and Tserkovnyak, Y.},
  title   = {Self-focusing skyrmion racetracks in ferrimagnets},
  journal = {Physical Review B},
  volume  = {95},
  pages   = {140404(R)},
  year    = {2017},
  doi     = {10.1103/PhysRevB.95.140404}
}

@article{Tveten2013,
  author  = {Tveten, E. G. and Qaiumzadeh, A. and Tretiakov, O. A. and Brataas, A.},
  title   = {Staggered Dynamics in Antiferromagnets by Collective Coordinates},
  journal = {Physical Review Letters},
  volume  = {110},
  pages   = {127208},
  year    = {2013},
  doi     = {10.1103/PhysRevLett.110.127208}
}

@article{Tveten2016,
  author  = {Tveten, E. G. and M{\"u}ller, T. and Linder, J. and Brataas, A.},
  title   = {Intrinsic magnetization of antiferromagnetic textures},
  journal = {Physical Review B},
  volume  = {93},
  pages   = {104408},
  year    = {2016},
  doi     = {10.1103/PhysRevB.93.104408}
}

@article{Nakane2021,
  author  = {Nakane, J. J. and Kohno, H.},
  title   = {Magnetic-Field-Driven Antiferromagnetic Domain Wall Motion},
  journal = {Journal of the Physical Society of Japan},
  volume  = {90},
  pages   = {034702},
  year    = {2021},
  doi     = {10.7566/JPSJ.90.034702}
}

@article{Hals2011,
  author  = {Hals, K. M. D. and Tserkovnyak, Y. and Brataas, A.},
  title   = {Phenomenology of current-induced dynamics in antiferromagnets},
  journal = {Physical Review Letters},
  volume  = {106},
  pages   = {107206},
  year    = {2011},
  doi     = {10.1103/PhysRevLett.106.107206}
}

@article{Gomonay2010,
  author  = {Gomonay, H. V. and Loktev, V. M.},
  title   = {Spin transfer and current-induced switching in antiferromagnets},
  journal = {Physical Review B},
  volume  = {81},
  pages   = {144427},
  year    = {2010},
  doi     = {10.1103/PhysRevB.81.144427}
}

@article{Oh2018,
  author  = {Oh, S.-H. and Lee, K.-J.},
  title   = {Ferrimagnetic domain wall motion induced by damping-like spin-orbit torque},
  journal = {Journal of Magnetics},
  volume  = {23},
  pages   = {196--200},
  year    = {2018},
  doi     = {10.4283/JMAG.2018.23.2.196}
}

@article{Oh2019,
  author  = {Oh, S.-H. and Kim, S. K. and Xiao, J. and Lee, K.-J.},
  title   = {Bidirectional spin-wave-driven domain wall motion in ferrimagnets},
  journal = {Physical Review B},
  volume  = {100},
  pages   = {174403},
  year    = {2019},
  doi     = {10.1103/PhysRevB.100.174403}
}

@article{Jing2022,
  author  = {Jing, K. Y. and Gong, X. and Wang, X. R.},
  title   = {Field-driven domain wall motion in ferrimagnetic nanowires},
  journal = {Physical Review B},
  volume  = {106},
  pages   = {174429},
  year    = {2022},
  doi     = {10.1103/PhysRevB.106.174429}
}

@article{Martinez2019,
  author  = {Martinez, E. and Raposo, V. and Alejos, O.},
  title   = {Current-driven domain wall dynamics in ferrimagnets: Micromagnetic approach and collective coordinates model},
  journal = {Journal of Magnetism and Magnetic Materials},
  volume  = {491},
  pages   = {165545},
  year    = {2019},
  doi     = {10.1016/j.jmmm.2019.165545}
}

@article{Yan2009,
  author  = {Yan, P. and Wang, X. R.},
  title   = {Domain wall propagation due to the synchronization with circularly polarized microwaves},
  journal = {Physical Review B},
  volume  = {80},
  pages   = {214426},
  year    = {2009},
  doi     = {10.1103/PhysRevB.80.214426}
}

@article{Mougin2007,
  author  = {Mougin, A. and Cormier, M. and Adam, J. P. and Metaxas, P. J. and Ferr{\'e}, J.},
  title   = {Domain wall mobility, stability and {W}alker breakdown in magnetic nanowires},
  journal = {Europhysics Letters},
  volume  = {78},
  pages   = {57007},
  year    = {2007},
  doi     = {10.1209/0295-5075/78/57007}
}

@article{Parkin2008,
  author  = {Parkin, S. S. P. and Hayashi, M. and Thomas, L.},
  title   = {Magnetic Domain-Wall Racetrack Memory},
  journal = {Science},
  volume  = {320},
  pages   = {190--194},
  year    = {2008},
  doi     = {10.1126/science.1145799}
}

@article{Parkin2015,
  author  = {Parkin, Stuart and Yang, See-Hun},
  title   = {Memory on the racetrack},
  journal = {Nature Nanotechnology},
  volume  = {10},
  pages   = {195--198},
  year    = {2015},
  doi     = {10.1038/nnano.2015.41}
}

@article{Yang2015,
  author  = {Yang, See-Hun and Ryu, Kyu-Sun and Parkin, Stuart},
  title   = {Domain-wall velocities of up to 750 m s$^{-1}$ driven by exchange-coupling torque in synthetic antiferromagnets},
  journal = {Nature Nanotechnology},
  volume  = {10},
  pages   = {221--226},
  year    = {2015},
  doi     = {10.1038/nnano.2014.324}
}

@article{Emori2013,
  author  = {Emori, S. and Bauer, U. and Ahn, S.-M. and Martinez, E. and Beach, G. S. D.},
  title   = {Current-driven dynamics of chiral ferromagnetic domain walls},
  journal = {Nature Materials},
  volume  = {12},
  pages   = {611--616},
  year    = {2013},
  doi     = {10.1038/nmat3675}
}

@article{Yang2021ChiralSpintronics,
  author  = {Yang, See-Hun and Naaman, Ron and Paltiel, Yossi and Parkin, Stuart S. P.},
  title   = {Chiral spintronics},
  journal = {Nature Reviews Physics},
  volume  = {3},
  pages   = {328--343},
  year    = {2021},
  doi     = {10.1038/s42254-021-00302-9}
}

@incollection{Alejos2020DWReview,
  author        = {Alejos, Oscar and Raposo, V{\'i}ctor and Mart{\'i}nez, Eduardo},
  title         = {Domain Wall Motion in Magnetic Nanostrips},
  booktitle     = {Materials Science and Technology},
  publisher     = {Wiley-VCH},
  address       = {Weinheim},
  pages         = {1--49},
  year          = {2020},
  doi           = {10.1002/9783527603978.mst0459}
}

@article{Tretiakov2008,
  author  = {Tretiakov, O. A. and Clarke, D. and Chern, G.-W. and Bazaliy, Ya. B. and Tchernyshyov, O.},
  title   = {Dynamics of Domain Walls in Magnetic Nanostrips},
  journal = {Physical Review Letters},
  volume  = {100},
  pages   = {127204},
  year    = {2008},
  doi     = {10.1103/PhysRevLett.100.127204}
}

@article{Baltz2018,
  author  = {Baltz, V. and Manchon, A. and Tsoi, M. and Moriyama, T. and Ono, T. and Tserkovnyak, Y.},
  title   = {Antiferromagnetic spintronics},
  journal = {Reviews of Modern Physics},
  volume  = {90},
  pages   = {015005},
  year    = {2018},
  doi     = {10.1103/RevModPhys.90.015005}
}

@article{Gomonay2016,
  author  = {Gomonay, O. and Jungwirth, T. and Sinova, J.},
  title   = {High antiferromagnetic domain wall velocity induced by {N}{\'e}el spin-orbit torques},
  journal = {Physical Review Letters},
  volume  = {117},
  pages   = {017202},
  year    = {2016},
  doi     = {10.1103/PhysRevLett.117.017202}
}

@article{Shiino16,
  author  = {Shiino, T. and Oh, S.-H. and Haney, P. M. and Lee, S.-W. and Go, G. and Park, B.-G. and Lee, K.-J.},
  title   = {Antiferromagnetic Domain Wall Motion Driven by Spin-Orbit Torques},
  journal = {Physical Review Letters},
  volume  = {117},
  pages   = {087203},
  year    = {2016},
  doi     = {10.1103/PhysRevLett.117.087203}
}

@article{Kim2022NatMater,
  author  = {Kim, S. and Beach, G. S. D. and Lee, K.-J. and Ono, T. and Rasing, T. and Yang, H.},
  title   = {Ferrimagnetic spintronics},
  journal = {Nature Materials},
  volume  = {21},
  pages   = {24--34},
  year    = {2022},
  doi     = {10.1038/s41563-021-01139-4}
}

@article{Kim2017NatMater,
  author  = {Kim, K.-J. and Kim, S. K. and Hirata, Y. and Oh, S.-H. and Tono, T. and Kim, D.-H. and Okuno, T. and Ham, W. S. and Kim, S. and Go, G. and Tsukamoto, A. and Tserkovnyak, Y. and Lee, K.-J. and Ono, T.},
  title   = {Fast domain wall motion in the vicinity of the angular momentum compensation temperature of ferrimagnets},
  journal = {Nature Materials},
  volume  = {16},
  pages   = {1187--1192},
  year    = {2017},
  doi     = {10.1038/nmat4990}
}

@article{Oh2017,
  author  = {Oh, S.-H. and Kim, S. K. and Lee, D.-K. and Go, G. and Kim, K.-J. and Ono, T. and Tserkovnyak, Y. and Lee, K.-J.},
  title   = {Coherent terahertz spin-wave emission associated with ferrimagnetic domain wall dynamics},
  journal = {Physical Review B},
  volume  = {96},
  pages   = {100407(R)},
  year    = {2017},
  doi     = {10.1103/PhysRevB.96.100407}
}

@article{Caretta20,
  author  = {Caretta, L. and Oh, S.-H. and Fakhrul, T. and Lee, D.-K. and Lee, B. H. and Kim, S. K. and Ross, C. A. and Lee, K.-J. and Beach, G. S. D.},
  title   = {Relativistic kinematics of a magnetic soliton},
  journal = {Science},
  volume  = {370},
  pages   = {1438--1442},
  year    = {2020},
  doi     = {10.1126/science.aba5555}
}

@article{Yurlov2021,
  author  = {Yurlov, V. V. and Zvezdin, K. A. and Skirdkov, P. N. and Zvezdin, A. K.},
  title   = {Domain wall dynamics of ferrimagnets influenced by spin current near the angular momentum compensation temperature},
  journal = {Physical Review B},
  volume  = {103},
  pages   = {134442},
  year    = {2021},
  doi     = {10.1103/PhysRevB.103.134442}
}

@article{Kim2020,
  author  = {Kim, D.-H. and Kim, D.-H. and Kim, K.-J. and Moon, K.-W. and Yang, S. and Lee, K.-J. and Kim, S. K.},
  title   = {The dynamics of a domain wall in ferrimagnets driven by spin-transfer torque},
  journal = {Journal of Magnetism and Magnetic Materials},
  volume  = {514},
  pages   = {167237},
  year    = {2020},
  doi     = {10.1016/j.jmmm.2020.167237}
}

@article{Kim2020PRB,
  author  = {Kim, D.-H. and Kim, D.-H. and Kim, D.-Y. and Choe, S.-B. and Ono, T. and Lee, K.-J. and Kim, S. K.},
  title   = {Magnetic soliton rectifier via phase synchronization},
  journal = {Physical Review B},
  volume  = {102},
  pages   = {184430},
  year    = {2020},
  doi     = {10.1103/PhysRevB.102.184430}
}

@article{Li2020,
  author  = {Li, W. H. and Jin, Z. and Wen, D. L. and Zhang, X. M. and Qin, M. H. and Liu, J.-M.},
  title   = {Ultrafast domain wall motion in ferrimagnets induced by magnetic anisotropy gradient},
  journal = {Physical Review B},
  volume  = {101},
  pages   = {024414},
  year    = {2020},
  doi     = {10.1103/PhysRevB.101.024414}
}

@article{Tejerina2020,
  author  = {S{\'a}nchez-Tejerina, L. and Puliafito, V. and {Khalili Amiri}, P. and Carpentieri, M. and Finocchio, G.},
  title   = {Dynamics of domain-wall motion driven by spin-orbit torque in antiferromagnets},
  journal = {Physical Review B},
  volume  = {101},
  pages   = {014433},
  year    = {2020},
  doi     = {10.1103/PhysRevB.101.014433}
}

@article{Pan2018,
  author  = {Pan, K. and Xing, L. and Yuan, H. Y. and Wang, W.},
  title   = {Driving chiral domain walls in antiferromagnets using rotating magnetic fields},
  journal = {Physical Review B},
  volume  = {97},
  pages   = {184418},
  year    = {2018},
  doi     = {10.1103/PhysRevB.97.184418}
}

@article{Jin2021,
  author  = {Jin, M. and Hong, I.-S. and Kim, D.-H. and Lee, K.-J. and Kim, S. K.},
  title   = {Domain-wall motion driven by a rotating field in a ferrimagnet},
  journal = {Physical Review B},
  volume  = {104},
  pages   = {184431},
  year    = {2021},
  doi     = {10.1103/PhysRevB.104.184431}
}

@article{Hardt2025,
  author  = {Hardt, D. and Doostani, R. and Diehl, S. and del Ser, N. and Rosch, A.},
  title   = {Propelling ferrimagnetic domain walls by dynamical frustration},
  journal = {Nature Communications},
  volume  = {16},
  pages   = {3829},
  year    = {2025},
  doi     = {10.1038/s41467-025-58920-1}
}

@article{Vansteenkiste2014,
  author = {Vansteenkiste, A. and Leliaert, J. and Dvornik, M. and Helsen, M. and Garcia-Sanchez, F. and Van Waeyenberge, B.},
  title = {The design and verification of MuMax3},
  journal = {AIP Advances},
  volume = {4},
  number = {10},
  pages = {107133},
  year = {2014},
  doi = {10.1063/1.4899186}
}

@article{Tatara2008,
  author = {Tatara, Gen and Kohno, Hiroshi and Shibata, Junya},
  title = {Microscopic approach to current-driven domain wall dynamics},
  journal = {Physics Reports},
  volume = {468},
  number = {6},
  pages = {213--301},
  year = {2008},
  doi = {10.1016/j.physrep.2008.07.003}
}

@book{Auerbach1994,
  author = {Auerbach, Assa},
  title = {Interacting Electrons and Quantum Magnetism},
  publisher = {Springer},
  year = {1994}
}

@article{Li2020JMMM,
  author  = {Li, W. H. and Chen, Z. Y. and Wen, D. L. and Chen, D. Y. and Fan, Z. and Zeng, M. and Lu, X. B. and Gao, X. S. and Qin, M. H.},
  title   = {Rotating magnetic field driven antiferromagnetic domain wall motion: Role of {Dzyaloshinskii-Moriya} interaction},
  journal = {Journal of Magnetism and Magnetic Materials},
  volume  = {497},
  pages   = {166051},
  year    = {2020},
  doi     = {10.1016/j.jmmm.2019.166051}
}

@article{Liu2022,
  author  = {Liu, T.-T. and Hu, Y.-F. and Liu, Y. and Jin, Z.-J.-Y. and Tang, Z.-H. and Qin, M.-H.},
  title   = {Domain wall dynamics driven by a circularly polarized magnetic field in ferrimagnet: effect of {Dzyaloshinskii--Moriya} interaction},
  journal = {Rare Metals},
  volume  = {41},
  number  = {11},
  pages   = {3815--3822},
  year    = {2022},
  doi     = {10.1007/s12598-022-02056-4}
}

@article{Chen2018,
  author  = {Chen, Z. Y. and Yan, Z. R. and Zhang, Y. L. and Qin, M. H. and Fan, Z. and Lu, X. B. and Gao, X. S. and Liu, J.-M.},
  title   = {Microwave fields driven domain wall motions in antiferromagnetic nanowires},
  journal = {New Journal of Physics},
  volume  = {20},
  pages   = {063003},
  year    = {2018},
  doi     = {10.1088/1367-2630/aac68e}
}

@article{Bisig2015,
  author  = {Bisig, A. and Mawass, M. A. and St{\"a}rk, M. and Moutafis, C. and Rhensius, J. and Heidler, J. and Gliga, S. and Weigand, M. and Tyliszczak, T. and Van Waeyenberge, B. and Stoll, H. and Sch{\"u}tz, G. and Kl{\"a}ui, M.},
  title   = {Dynamic domain wall chirality rectification by rotating magnetic fields},
  journal = {Applied Physics Letters},
  volume  = {106},
  pages   = {122401},
  year    = {2015},
  doi     = {10.1063/1.4915256}
}

@article{Mulkers2017,
  author  = {Mulkers, J. and Van Waeyenberge, B. and Milo{\v{s}}evi{\'{c}}, M. V.},
  title   = {Effects of spatially engineered {D}zyaloshinskii-{M}oriya interaction in ferromagnetic films},
  journal = {Physical Review B},
  volume  = {95},
  number  = {14},
  pages   = {144401},
  year    = {2017},
  doi     = {10.1103/PhysRevB.95.144401}
}

@article{Thiaville2012,
  author  = {Thiaville, A. and Rohart, S. and Ju{\'e}, {\'E}. and Cros, V. and Fert, A.},
  title   = {Dynamics of {D}zyaloshinskii domain walls in ultrathin magnetic films},
  journal = {Europhysics Letters},
  volume  = {100},
  pages   = {57002},
  year    = {2012},
  doi     = {10.1209/0295-5075/100/57002}
}

@article{Krishnia2021,
  author  = {Krishnia, S. and Haltz, E. and Berges, L. and Aballe, L. and Foerster, M. and Bocher, L. and Weil, R. and Thiaville, A. and Sampaio, J. and Mougin, A.},
  title   = {Spin-orbit coupling in single-layer ferrimagnets: Direct observation of spin-orbit torques and chiral spin textures},
  journal = {Physical Review Applied},
  volume  = {16},
  pages   = {024040},
  year    = {2021},
  doi     = {10.1103/PhysRevApplied.16.024040}
}

@book{LLMechanics,
  author    = {Landau, L. D. and Lifshitz, E. M.},
  title     = {Mechanics},
  edition   = {3rd},
  series    = {Course of Theoretical Physics},
  volume    = {1},
  publisher = {Pergamon},
  address   = {Oxford},
  year      = {1976},
  note      = {{Sec.}~39}
}

\end{document}